\documentstyle[epsfig,12pt]{article}
\begin{document}
\title{
\rightline{\small  UL--NTZ 28/97}
Helicity amplitudes for the small angle lepton pair production in
$e^+e^-$ or $\mu^+ \mu^-$ collisions }

\author{E.A.~Kuraev$^1$\thanks{E-mail:
kuraev@thsun1.jinr.dubna.su},
A.~Schiller$^2$\thanks{E-mail: schiller@tph204.physik.uni-leipzig.de},
V.G.~Serbo$^3$\thanks{E-mail: serbo@math.nsc.ru}
and D.V.~Serebryakova$^4$\\
$^1$ Joint Institute of Nuclear Research, 141980, Dubna, Russia\\
$^2$ Institut f\"ur Theoretische Physik and
Naturw.-Theoretisches \\ Zentrum,
Universit\"at Leipzig,  D-04109 Leipzig, Germany\\
$^3$ Novosibirsk State University, 630090, Novosibirsk, Russia\\
$^4$ Institute of Mathematics, 630090, Novosibirsk, Russia }

\date{October 21, 1997}

\maketitle

\begin{abstract}

The lepton pair production $e^- e^+ \to e^- e^+ l^-  l^+ $ or
$\mu^- \mu^+ \to \mu^- \mu^+ l^- l^+ $ is
studied in
the dominant cross section
region of scattering angles ${m_j}/{E_j} \stackrel{<}{\sim}
\theta_j  \ll 1 $. An analytical expression is found for all 64
helicity amplitudes of these processes. The accuracy of
the obtained formulae is given omitting only terms of the
order of ${m_j^2}/{E_j^2}$, $\theta_j^2$ and $ \theta_j
{m_j}/{E_j}$. The result has a compact form convenient both for
analytical and numerical calculations of various cross
sections in this dominating scattering regime.
\end{abstract}

\section{Introduction}
\label{sec:1}
Colliders with electron and photon beams are now widely used or
designed to study fundamental interactions. Among other QED
reactions those inelastic processes are of special interest the
cross section of which do not drop with increasing energies. To
third and fourth orders in the electromagnetic coupling $\alpha$
these processes are shown in Figs.~\ref{fig:1}-\ref{fig:8} (only
block diagrams are presented).
\begin{figure}[!htb]
  \centering
  \setlength{\unitlength}{1cm}
  \begin{minipage}[t]{7cm}
  \begin{picture}(7.0,3.0)
\unitlength=2.00mm
\begin{picture}(24.00,14.00)
\put(12.00,12.00){\circle{3.40}}
\put( 0.00, 0.00){\line(1,0){12.00}}
\put( 0.00, 0.00){\vector(1,0){6.4}} 
\put( 0.00,12.00){\line(1,0){2.0}}
\put( 2.50,12.00){\line(1,0){2.0}}
\put( 5.00,12.00){\line(1,0){2.0}}
\put( 5.00,12.00){\vector(1,0){1.40}}
\put( 7.50,12.00){\line(1,0){2.0}}
\put(10.00,12.00){\line(1,0){0.2}}
\put(12.00, 0.00){\line(0,1){2.0}}
\put(12.00, 2.50){\line(0,1){2.0}}
\put(12.00, 5.00){\line(0,1){2.0}}
\put(12.00, 5.00){\vector(0,1){1.40}}
\put(12.00, 7.50){\line(0,1){2.0}}
\put(12.00,10.00){\line(0,1){0.2}}
\put(12.00, 0.00){\vector(1,0){6.4}} 
\put(12.00, 0.00){\line(1,0){12.0}}
\put(13.40,10.90){\line(1,0){10.6}}
\put(13.40,13.10){\vector(1,0) {5.00}}
\put(24.00,10.90){\vector(-1,0){6.40}}
\put(13.40,13.10){\line(1,0){10.60}}
\put(18.50,14.60){\makebox(0,0)[cc]{$p_-(E_-)$}}
\put(18.00, 8.90){\makebox(0,0)[cc]{$-p_+(E_+)$}}
\put(18.00, 1.30){\makebox(0,0)[cc]{$p_4(E_4)$}}
\put( 6.00,10.00){\makebox(0,0)[cc]{$k(\omega)$}}
\put( 6.00, 1.50){\makebox(0,0)[cc]{$p_2(E_2)$}}
\put(10.50, 5.50){\makebox(0,0)[cc]{$q$}}

\end{picture}
  \end{picture} \par
  \caption{ Reaction  $\gamma e \to l^+ l^- e $ }
  \label{fig:1}
  \end{minipage}\hfill
  \begin{minipage}[t]{7cm}
  \begin{picture}(7.0,3.0)
\unitlength=2.00mm
\begin{picture}(24.00,14.00)
\put(12.00, 0.00){\circle{3.40}}
\put(12.00,12.00){\circle{3.40}}
\put( 0.00, 0.00){\line(1,0){2.0}}
\put( 2.50, 0.00){\line(1,0){2.0}}
\put( 5.00, 0.00){\line(1,0){2.0}}
\put( 5.00, 0.00){\vector(1,0){1.40}}
\put( 7.50, 0.00){\line(1,0){2.0}}
\put(10.00, 0.00){\line(1,0){0.2}}
\put( 0.00,12.00){\line(1,0){2.0}}
\put( 2.50,12.00){\line(1,0){2.0}}
\put( 5.00,12.00){\line(1,0){2.0}}
\put( 5.00,12.00){\vector(1,0){1.40}}
\put( 7.50,12.00){\line(1,0){2.0}}
\put(10.00,12.00){\line(1,0){0.2}}
\put(12.00, 1.70){\line(0,1){0.3}}
\put(12.00, 2.50){\line(0,1){2.0}}
\put(12.00, 5.00){\line(0,1){2.0}}
\put(12.00, 5.00){\vector(0,1){1.40}}
\put(12.00, 7.50){\line(0,1){2.0}}
\put(12.00,10.00){\line(0,1){0.2}}
\put(13.40,-1.10){\line(1,0){10.6}} 
\put(24.00,-1.10){\vector(-1,0){6.40}}
\put(13.40, 1.10){\vector( 1,0){5.00}} 
\put(13.40, 1.10){\line(1,0){10.60}}
\put(13.40,10.90){\line(1,0){10.6}} 
\put(24.00,10.90){\vector(-1,0){6.40}}
\put(13.40,13.10){\vector( 1,0){5.00}} 
\put(13.40,13.10){\line(1,0){10.60}}
\end{picture}
  \end{picture} \par
  \caption{ Reaction $\gamma \gamma \to e^+ e^- l^+ l^- $ }
  \label{fig:2}
  \end{minipage}
\end{figure}
Figs.~\ref{fig:1} and \ref{fig:2} describe the lepton pair
production in $\gamma e$ and $\gamma \gamma$ collisions arising
from $\gamma \gamma^*$ interactions.
\begin{figure}[!htb]
  \centering
  \setlength{\unitlength}{1cm}
  \begin{minipage}[t]{7cm}
  \begin{picture}(7.0,3.0)
\unitlength=2.00mm
\begin{picture}(24.00,14.00)
\put(12.00,12.00){\circle{3.40}}
\put( 0.00, 0.00){\line(1,0){12.00}}
\put( 0.00, 0.00){\vector(1,0){6.4}} 
\put( 0.00,10.90){\line(1,0){10.60}}
\put( 0.00,10.90){\vector(1,0){6.4}} 
\put(12.00, 0.00){\line(0,1){2.0}}
\put(12.00, 2.50){\line(0,1){2.0}}
\put(12.00, 5.00){\line(0,1){2.0}}
\put(12.00, 5.00){\vector(0,1){1.40}}
\put(12.00, 7.50){\line(0,1){2.0}}
\put(12.00,10.00){\line(0,1){0.2}}
\put(12.00, 0.00){\vector(1,0){6.4}} 
\put(12.00, 0.00){\line(1,0){12.0}}
\put(13.40,13.10){\line(1,0){0.6}}
\put(14.50,13.10){\line(1,0){2.0}}
\put(17.00,13.10){\line(1,0){2.0}}
\put(17.00,13.10){\vector(1,0){1.40}}
\put(19.50,13.10){\line(1,0){2.0}}
\put(22.00,13.10){\line(1,0){2.0}}
\put(13.40,10.90){\vector(1,0){5.0}} 
\put(13.40,10.90){\line(1,0){10.6}}
\put(18.00,14.60){\makebox(0,0)[cc]{$k(\omega)$}}
\put(18.00, 8.90){\makebox(0,0)[cc]{$p_3(E_3)$}}
\put(18.00, 1.50){\makebox(0,0)[cc]{$p_4(E_4)$}}
\put( 6.00, 8.90){\makebox(0,0)[cc]{$p_1(E_1)$}}
\put( 6.00, 1.50){\makebox(0,0)[cc]{$p_2(E_2)$}}
\put(10.50, 5.50){\makebox(0,0)[cc]{$q$}}

\end{picture}
  \end{picture} \par
  \caption{Single bremsstrahlung $e e \to e e \gamma$}
  \label{fig:3}
  \end{minipage}\hfill
  \begin{minipage}[t]{7cm}
  \begin{picture}(7.0,3.0)
\unitlength=2.00mm
\begin{picture}(24.00,14.00)
\put(12.00, 0.00){\circle{3.40}}
\put(12.00,12.00){\circle{3.40}}
\put( 0.00, 1.10){\line(1,0){10.60}}
\put( 0.00, 1.10){\vector(1,0){6.4}} 
\put( 0.00,10.90){\line(1,0){10.60}}
\put( 0.00,10.90){\vector(1,0){6.4}} 
\put(12.00, 1.70){\line(0,1){0.2}}
\put(12.00, 2.50){\line(0,1){2.0}}
\put(12.00, 5.00){\line(0,1){2.0}}
\put(12.00, 5.00){\vector(0,1){1.40}}
\put(12.00, 7.50){\line(0,1){2.0}}
\put(12.00,10.00){\line(0,1){0.2}}
\put(13.40,-1.10){\line(1,0){0.6}}
\put(14.50,-1.10){\line(1,0){2.0}}
\put(17.00,-1.10){\line(1,0){2.0}}
\put(17.00,-1.10){\vector(1,0){1.40}}
\put(19.50,-1.10){\line(1,0){2.0}}
\put(22.00,-1.10){\line(1,0){2.0}}
\put(13.40, 1.10){\vector(1,0){5.0}} 
\put(13.40, 1.10){\line(1,0){10.6}}
\put(13.40,13.10){\line(1,0){0.6}}
\put(14.50,13.10){\line(1,0){2.0}}
\put(17.00,13.10){\line(1,0){2.0}}
\put(17.00,13.10){\vector(1,0){1.40}}
\put(19.50,13.10){\line(1,0){2.0}}
\put(22.00,13.10){\line(1,0){2.0}}
\put(13.40,10.90){\vector(1,0){5.0}} 
\put(13.40,10.90){\line(1,0){10.6}}
\end{picture}
  \end{picture} \par
  \caption{Double bremsstrahlung $e e \to e e \gamma \gamma$}
  \label{fig:4}
  \end{minipage}
\end{figure}
\begin{figure}[!htb]
  \centering
  \setlength{\unitlength}{1cm}
  \begin{minipage}[t]{7cm}
  \begin{picture}(7.0,3.0)
\unitlength=2.00mm
\begin{picture}(24.00,14.00)
\put(12.00, 6.00){\circle{3.40}}
\put(13.40, 4.90){\line(1,0){10.6}}
\put(13.40, 7.10){\vector(1,0) {5.00}}
\put(24.00, 4.90){\vector(-1,0){6.40}}
\put(13.40, 7.10){\line(1,0){10.60}}
\put( 0.00, 0.00){\line(1,0){12.00}}
\put(12.00, 0.00){\vector(-1,0){6.4}} 
\put( 0.00,12.00){\line(1,0){12.00}}
\put( 0.00,12.00){\vector(1,0){6.4}} 
\put(12.00, 0.00){\line(0,1){2.0}}
\put(12.00, 0.00){\vector(0,1){1.40}}
\put(12.00, 2.50){\line(0,1){1.7}}
\put(12.00, 7.80){\line(0,1){1.7}}
\put(12.00,10.00){\line(0,1){2.0}}
\put(12.00,12.00){\vector(0,-1){1.40}}
\put(24.00, 0.00){\vector(-1,0){6.4}} 
\put(12.00, 0.00){\line(1,0){12.0}}
\put(12.00,12.00){\vector(1,0){6.4}} 
\put(12.00,12.00){\line(1,0){12.0}}
\put( 6.00,13.50){\makebox(0,0)[cc]{$p_1$}}
\put( 6.00, 1.50){\makebox(0,0)[cc]{$-p_2$}}
\put(18.00,13.50){\makebox(0,0)[cc]{$p_3$}}
\put(18.00, 1.50){\makebox(0,0)[cc]{$-p_4$}}
\put(10.50, 9.50){\makebox(0,0)[cc]{$k$}}
\put(10.50, 2.50){\makebox(0,0)[cc]{$q$}}
\put(22.00, 9.00){\makebox(0,0)[cc]{$p_-$}}
\put(22.00, 3.00){\makebox(0,0)[cc]{$-p_+$}}

\end{picture}
  \end{picture} \par
  \caption{ Two--photon pair production $e^- e^+ \to e^- e^+ l^-
  l^+ $ }
  \label{fig:5}
  \end{minipage}\hfill
  \begin{minipage}[t]{7cm}
  \begin{picture}(7.0,3.0)
\unitlength=2.00mm
\begin{picture}(24.00,14.00)
\put(12.00,12.00){\circle{3.40}}
\put( 0.00, 0.00){\line(1,0){12.00}}
\put(12.00, 0.00){\vector(-1,0){6.4}} 
\put( 0.00,10.90){\line(1,0){10.60}}
\put( 0.00,10.90){\vector(1,0){6.4}} 
\put(12.00, 0.00){\line(0,1){2.0}}
\put(12.00, 2.50){\line(0,1){2.0}}
\put(12.00, 5.00){\line(0,1){2.0}}
\put(12.00, 5.00){\vector(0,1){1.40}}
\put(12.00, 7.50){\line(0,1){2.0}}
\put(12.00,10.00){\line(0,1){0.2}}
\put(24.00, 0.00){\vector(-1,0){6.4}} 
\put(12.00, 0.00){\line(1,0){12.0}}
\put(13.40,13.10){\line(1,0){0.6}}
\put(14.50,13.10){\line(1,0){2.0}}
\put(17.00,13.10){\line(1,0){2.0}}
\put(14.50,13.10){\vector(1,0){1.40}}
\put(19.00,13.10){\line(4,1){5.0}}
\put(19.00,13.10){\vector(4,1){2.9}}
\put(19.00,13.10){\line(4,-1){5.0}}
\put(24.00,11.85){\vector(-4, 1){2.9}}
\put(13.40,10.90){\vector(1,0){5.0}} 
\put(13.40,10.90){\line(1,0){10.6}}

\put(6.00,8.90){\makebox(0,0)[cc]{$p_1$}}
\put(6.00,1.50){\makebox(0,0)[cc]{$-p_2$}}
\put(18.00,8.90){\makebox(0,0)[cc]{$p_3$}}
\put(18.00,1.50){\makebox(0,0)[cc]{$-p_4$}}
\put(10.50,5.50){\makebox(0,0)[cc]{$q$}}
\put(16.00,14.60){\makebox(0,0)[cc]{$l$}}
\put(26.00,14.50){\makebox(0,0)[cc]{$p_-$}}
\put(26.00,12.00){\makebox(0,0)[cc]{$-p_+$}}
\end{picture}
  \end{picture} \par
  \caption{ Bremsstrahlung pair production $e^- e^+ \to e^- e^+
  l^- l^+ $ }
  \label{fig:6}
  \end{minipage}
\end{figure}
Figs.~\ref{fig:3} and \ref{fig:4} correspond to single and double
bremsstrahlung (with a single photon along its parent lepton),
Fig.~\ref{fig:5} and \ref{fig:6} to the lepton pair production
by the two--photon and bremsstrahlung mechanisms,
Fig.~\ref{fig:7} gives the process $\gamma e \to l^+l^- e \gamma$
and Fig.~\ref{fig:8} the double bremsstrahlung along one
direction.
\begin{figure}[!htb]
  \centering
  \setlength{\unitlength}{1cm}
  \begin{minipage}[t]{7cm}
  \begin{picture}(7.0,3.0)
\unitlength=2.00mm
\begin{picture}(24.00,14.00)
\put(12.00, 0.00){\circle{3.40}}
\put(12.00,12.00){\circle{3.40}}
\put( 0.00, 1.10){\line(1,0){10.60}}
\put( 0.00, 1.10){\vector(1,0){6.4}}
\put( 0.00,12.00){\line(1,0){2.0}}
\put( 2.50,12.00){\line(1,0){2.0}}
\put( 5.00,12.00){\line(1,0){2.0}}
\put( 5.00,12.00){\vector(1,0){1.40}}
\put( 7.50,12.00){\line(1,0){2.0}}
\put(10.00,12.00){\line(1,0){0.2}}
\put(12.00, 1.70){\line(0,1){0.2}}
\put(12.00, 2.50){\line(0,1){2.0}}
\put(12.00, 5.00){\line(0,1){2.0}}
\put(12.00, 5.00){\vector(0,1){1.40}}
\put(12.00, 7.50){\line(0,1){2.0}}
\put(12.00,10.00){\line(0,1){0.2}}
\put(13.40,-1.10){\line(1,0){0.6}}
\put(14.50,-1.10){\line(1,0){2.0}}
\put(17.00,-1.10){\line(1,0){2.0}}
\put(17.00,-1.10){\vector(1,0){1.40}}
\put(19.50,-1.10){\line(1,0){2.0}}
\put(22.00,-1.10){\line(1,0){2.0}}
\put(13.40, 1.10){\vector(1,0){5.0}}
\put(13.40, 1.10){\line(1,0){10.6}}
\put(13.40,10.90){\line(1,0){10.6}}
\put(13.40,13.10){\vector(1,0) {5.00}}
\put(24.00,10.90){\vector(-1,0){6.40}}
\put(13.40,13.10){\line(1,0){10.60}}

\end{picture}
  \end{picture} \par
  \caption{ Reaction  $\gamma e \to l^+ l^- e \gamma$ }
  \label{fig:7}
  \end{minipage}\hfill
  \begin{minipage}[t]{7cm}
  \begin{picture}(7.0,3.0)
\unitlength=2.00mm
\begin{picture}(24.00,14.00)
\put(12.00,12.00){\circle{3.40}}
\put( 0.00, 0.00){\line(1,0){12.00}}
\put( 0.00, 0.00){\vector(1,0){6.4}} 
\put( 0.00,10.90){\line(1,0){10.60}}
\put( 0.00,10.90){\vector(1,0){6.4}} 
\put(12.00, 0.00){\line(0,1){2.0}}
\put(12.00, 2.50){\line(0,1){2.0}}
\put(12.00, 5.00){\line(0,1){2.0}}
\put(12.00, 5.00){\vector(0,1){1.40}}
\put(12.00, 7.50){\line(0,1){2.0}}
\put(12.00,10.00){\line(0,1){0.2}}
\put(12.00, 0.00){\vector(1,0){6.4}} 
\put(12.00, 0.00){\line(1,0){12.0}}
\put(13.40,13.10){\line(1,0){0.6}}
\put(14.50,13.10){\line(1,0){2.0}}
\put(17.00,13.10){\line(1,0){2.0}}
\put(17.00,13.10){\vector(1,0){1.40}}
\put(19.50,13.10){\line(1,0){2.0}}
\put(22.00,13.10){\line(1,0){2.0}}
\put(13.70,12.00){\line(1,0){0.2}}
\put(14.50,12.00){\line(1,0){2.0}}
\put(17.00,12.00){\line(1,0){2.0}}
\put(17.00,12.00){\vector(1,0){1.40}}
\put(19.50,12.00){\line(1,0){2.0}}
\put(22.00,12.00){\line(1,0){2.0}}

\put(13.40,10.90){\vector(1,0){5.0}} 
\put(13.40,10.90){\line(1,0){10.6}}
\end{picture}
  \end{picture} \par
  \caption{ Double
  bremsstrahlung  $e e \to e e \gamma \gamma$  in one direction}
  \label{fig:8}
  \end{minipage}
\end{figure}

The described processes are important by the following reasons

{\it (i)} Some of these reactions are used (or are proposed to
be used) as the monitoring processes to determine the collider
luminosity and to measure the polarisation of the colliding
particles. For example, the double bremsstrahlung
Fig.~\ref{fig:2} has been used as the standard calibration
process at several colliders in Novosibirsk, Frascati and Orsay
\cite{baier}-\cite{augustin}.  In \cite{leplumsag} it has been
suggested to use the single bremsstrahlung Fig.~\ref{fig:1} for
measuring the luminosity and the polarisation
of the initial $e^\pm$ at the LEP collider (see also the
paper~\cite{kotkinpl89}). It has been demonstrated in an
experiment \cite{leplum} that  this single bremsstrahlung process
has a  good chance to be used for luminosity purposes.
Recently~\cite{courau} the same process is proposed to measure
the  luminosity  at the DA$\Phi$NE collider. The processes
$\gamma \gamma \to \mu^+ \mu^- e^+ e^-$ and $\gamma \gamma \to
\mu^+ \mu^- \mu^+ \mu^-$ may be useful to monitor  colliding
$\gamma \gamma$ beams \cite{beams,kuraevpl84,kuraevnp85}.
Finally, the possibility of designing $\mu^+ \mu^-$ colliders is
widely discussed \cite{mumucoll} at present. Therefore, the
processes $\mu^+ \mu^- \to l^+ l^- l^+l^- $ ($l=e,\mu$) may be
useful for luminosity measurements at those
colliders~\cite{mumuGinz}.

{\it (ii)} Due to their large cross sections those reactions
contribute as a  significant background to a number of
experiments in the electroweak sector and to hadronic cross
sections. For example, the background process $e^+e^- \to e^+ e^-
\mu^+ \mu^-$ is of special importance for experiments studying
the two--photon $\pi^+\pi^-$ production due to the known
experimental difficulties in discriminating pions and muons.

{\it (iii)} The methods to calculate the helicity amplitudes
of those processes
 and to obtain some  distributions for them can be easily
translated to several  semihard QCD processes such as $\gamma
\gamma \to q \overline{q} Q \overline{Q}$ \cite{kuraevnp85} ($q$
and $Q$ are different quarks) and $\gamma \gamma \to M M'$,
$\gamma \gamma \to M q \overline{q}$ \cite{mesons} ($M, M'$
denote neutral mesons  as $\rho^0,\, \omega,\, \phi,\, \Psi,\,
\pi^0,\, a_2 ...$).

At high energies with the condition ($m_j$ are the lepton masses)
\begin{equation}
s= 2 p_1 p_2 = 4 E_1 E_2 \gg m_j^2
\label{1}
\end{equation}
the dominant contribution to the cross sections of
Figs.~\ref{fig:1}-\ref{fig:8}
are given by the region of scattering angles $\theta_j$ which are
much smaller than unity though they may be of the order of the
typical emission angles $m_j / E_j$ or larger:
\begin{equation}
\frac{m_j}{E_j} \stackrel{<}{\sim} \theta_j  \ll 1 \ .
\label{2}
\end{equation}
In this region all processes have the form of two--jet
processes with an exchange of a single virtual photon $\gamma^*$
 in the $t$--channel (see
Fig.~\ref{fig:9}).
\begin{figure}[!htb]
  \centering
\unitlength=2.0mm
\begin{picture}(38.00,14.00)
\put(23.00,12.00){\circle{4.00}}
\put(23.00,3.00){\circle{4.00}}
\put(9.00,3.00){\vector(1,0){6.50}}
\put(9.00,12.00){\vector(1,0){6.50}}
\put(14.00,3.00){\line(1,0){6.90}}
\put(14.00,12.00){\line(1,0){6.90}}
\put(25.00,12.50){\vector(1,0){6.00}}
\put(25.00,11.50){\vector(1,0){6.00}}
\put(24.50,13.50){\vector(1,0){6.50}}
\put(23.00,5.10){\line(0,1){0.90}}
\put(23.00,6.50){\line(0,1){2.00}}
\put(23.00,6.50){\vector(0,1){1.40}}
\put(23.00,9.00){\line(0,1){0.90}}
\put(24.50,10.50){\vector(1,0){6.50}}
\put(25.00,3.50){\vector(1,0){6.00}}
\put(25.00,2.50){\vector(1,0){6.00}}
\put(24.50,4.50){\vector(1,0){6.50}}
\put(24.50,1.50){\vector(1,0){6.50}}
\put(33.00,12.00){\makebox(0,0)[cc]{$p_j$}}
\put(21.00,7.00){\makebox(0,0)[cc]{$q$}}
\put(5.00,12.00){\makebox(0,0)[cc]{$p_1(E_1)$}}
\put(5.00,3.00){\makebox(0,0)[cc]{$p_2(E_2)$}}
\put(38.00,12.00){\makebox(0,0)[cc]{{\Large\}}$jet_1$}}
\put(38.00,3.00){\makebox(0,0)[cc]{{\Large\}}$jet_2$}}
\end{picture}
  \caption{ Generic block diagram  $e e \to  {\mathrm {jet}}_1 \
  {\mathrm{jet}}_2$ }
  \label{fig:9}
\end{figure}

The considered QED processes are widely discussed in the
literature. Taking into account unpolarised leptons and photons
only, various differential cross sections are summarised in
reviews \cite{bgms} and \cite{bfkk}. Recently the considered
reactions have been taken into account as radiative corrections
to the unpolarised Bhabha scattering used as calibration
process at LEP~\cite{recent}.

In the nearest future beams with polarised leptons and photons
will be available which demands to calculate cross sections with
polarised particles. In this connection we would like to note
that in the kinematic region (\ref{1}-\ref{2}) the discussed
QED processes can be found in a ``final form'' including the
polarisations of all particles. By this we mean that it is
possible to obtain compact and simple analytical expressions for
all helicity amplitudes with high accuracy. Omitting terms of
the order of
\begin{equation}
\frac{m_j^2}{E_j^2} \ , \ \ \theta_j^2 \ , \ \ \frac{m_j}{E_j}
\theta_j
\label{3}
\end{equation}
only, the amplitude $M_{fi}$ of any process given in
Figs.~\ref{fig:1}-\ref{fig:8} can be represented in a simple
factorised form
\begin{equation}
M_{fi}= \frac{s}{q^2} J_1 J_2 \ .
\label{4}
\end{equation}
The vertex factor $J_1$ ($J_2$)  corresponds to the first jet
or upper block (second jet or lower block) of Fig.~\ref{fig:9}.
The factor $J_{1(2)}$  depends on the energy fraction
$x_j=E_j/E_{1(2)}$, the transverse momenta ${\bf p}_{j\perp}$,
the helicities $\lambda_i$ of the particles in the first
(second) jet and the helicity $\lambda_{1(2)}$ of the initial
particle with 4--momentum $p_{1(2)}$. The function $J_{1(2)}$ is
independent on the cms energy squared $s$.

This approximation  differs considerably from the known results
of the CALCUL group and others \cite{CALCUL,kuraevCALCUL} where
such processes are calculated for not too small scattering angles
of the final particles. This allows to neglect completely lepton
masses, i.e. neglecting the terms of the order of $m_j/|{\bf
p}_{j\perp}|$, which, however, give the dominant contribution to
the total cross section at small angles.

For the reactions of Figs.~\ref{fig:1}-\ref{fig:4},\ref{fig:7}
the corresponding vertex factors and some differential cross
sections have been found
in~\cite{kuraevpl84,kuraevnp85,kuraevzp86}. These vertex
factors are presented in Sec.~\ref{sec:2} and partly used below, too.

In the present paper we give a complete set of helicity
amplitudes for the lepton pair production at $e^\pm e^-$ and
$\mu^{\pm}\mu^-$ colliders (see Figs.~\ref{fig:5} and
\ref{fig:6}) in the region (\ref{1}-\ref{2}). For
definiteness, we consider the process
\begin{equation}
e^- e^+ \to e^- e^+ \mu^- \mu^+ \ .
\label{5}
\end{equation}
In Sec.~\ref{sec:3} we derive the vertex factors necessary to calculate
the transition
amplitude of the two--photon lepton pair production of Fig.~\ref{fig:5}.
The following Section is devoted to the pair production via bremsstrahlung.
In Sec.~\ref{sec:5} our results are briefly summarised and
qualitative features of cross sections are discussed.

To complete the jet--like QED tree processes to fourth order,
the double brems\-strahlung along one jet direction has to be
calculated for the kinematic region under discussion. This work
is now in progress.

Let us introduce some notations using the block diagram of
Fig.~\ref{fig:9} as an example. We use a reference frame in which
the initial particles with 4--momenta $p_1$ and $p_2$ perform a
head--on collision with energies $E_1$ and $E_2$ of the same
order (e.g. at the B--factory). The $z$--axis is chosen along
the momentum ${\bf p}_1$, the azimuthal angles are denoted by
$\varphi_i$ (they are referred to one fixed $x$-axis). It
is convenient to introduce ``the almost light--like'' 4--vectors
$p$ and $p'$
$$
p=p_1 -{m^2\over s}\, p_2, \;
p'=p_2 -{m^2\over s}\, p_1, \;
p^2 ={p'}^2 = {m^6\over s^2},
$$
\begin{equation}
s= 2p_1 p_2 = 2 p p' + {3m^4\over s} \, .
\label{6}
\end{equation}
Throughout the paper we use the  notation ${\bf Q}$ and $R$ which
enter in the definition of the vertex factors
\begin{equation}
{\bf Q} = {{\bf u}\over a}+ {{\bf v}\over b}, \;\;\;
R       = {1\over a} - {1\over b}\,, \;\;\;
{\bf u}+ {\bf v} = {\bf q}_\perp\, , \;\;\;
b-a= {\bf v}^2 - {\bf u}^2\, .
\label{7}
\end{equation}
${\bf q}_\perp$ denotes the transverse momentum of the t--channel
virtual photon.
The quantities ${\bf u}$, ${\bf v}$, $a$ and $b$ are process
dependent.
Note the following useful relation between ${\bf Q}^2$ and $R^2$
\begin{equation}
{\bf Q}^2 + (a-{\bf u}^2) R^2 = \frac{{\bf q}_\perp^2}{ab} \, .
\label{7a}
\end{equation}
 Additionally, the helicity vectors
\begin{equation}
{\bf e}_\lambda = - \frac{\lambda}{\sqrt 2} ( 1,i\lambda,0)=
- {\bf e}^*_{-\lambda}, \ \ \lambda=\pm 1 \
\label{8}
\end{equation}
are used to describe the photon polarisation.
\section{A short description of our previous results}
\label{sec:2}

The amplitude $M_{fi}$ corresponding to the diagram of
Fig.~\ref{fig:9} can be presented in the form
\begin{equation}
M_{fi} = M_1^{\mu} \; {g_{\mu \nu} \over q^2} \; M_2^\nu ,
\label{9}
\end{equation}
where $M_1^\mu$ and $M_2^\nu$ are the amplitudes of the upper and
lower block of Fig.~\ref{fig:9}, respectively,  $(-g_{\mu \nu}/2)$ denotes the
density matrix of the virtual photon. The transition amplitude
$M_1$ describes the scattering of an incoming lepton or photon
with a virtual photon of ``mass'' squared $q^2$ and
polarisation vector ${\bf e}={\bf q}_\perp/|{\bf q}_\perp|$ to
some QED final state in the jet kinematics
(\ref{1}-\ref{2})
(similar for $M_2$).

With accuracy (\ref{3}) the $g_{\mu\nu}$ matrix can be
transformed to the form
\begin{equation}
g_{\mu \nu} \; \to {2\over s} \, p'_\mu\, p_\nu
\label{10}
\end{equation}
(for detail see \cite{ABgreen}, \S 4.8.4) which results in Eq.~(\ref{4}).
The vertex factors $J_{1,2}$ are given by the block amplitudes $M_{1,2}^\mu$
\begin{equation}
M_{fi} = {s\over q^2}\, J_1 \, J_2 , \;\;\;
J_1 = {\sqrt{2}\over s} \, M_1^{\mu} \, p'_\mu ,\; \;\;
J_2 = {\sqrt{2}\over s} \, M_2^{\mu} \, p_\mu \, .
\label{11}
\end{equation}
The quantities $J_1$ and $J_2$ can be calculated in the limit
$s \to \infty$ assuming that the  energy fractions and transverse
momenta of the final particles ${\bf p}_{i \perp}$ are finite in
this limit.
For the convenience of the reader we present in the following
the vertex factors corresponding to processes of
Figs.~\ref{fig:1}-\ref{fig:4} taken from
Refs.~\cite{kuraevnp85,kuraevzp86}.

\begin{figure}[!htb]
  \centering
\unitlength=2.00mm
\begin{picture}(24.00,12.00)(0.00,5.00)
\put( 0.00,12.00){\line(1,0){12.00}}
\put( 0.00,12.00){\vector(1,0){6.4}} 
\put(12.00, 5.00){\line(0,1){2.0}}
\put(12.00, 7.50){\line(0,1){2.0}}
\put(12.00, 7.50){\vector(0,1){1.40}}
\put(12.00,10.00){\line(0,1){2.0}}
\put(12.00,12.00){\vector(1,0){6.4}} 
\put(12.00,12.00){\line(1,0){12.0}}

\put( 6.00,10.50){\makebox(0,0)[cc]{$p_1$}}
\put(18.00,10.50){\makebox(0,0)[cc]{$p_3$}}
\put(10.50, 8.00){\makebox(0,0)[cc]{$q$}}
\end{picture}
  \caption{ Amplitude $e^- \gamma^* \to e^-$ for
  $J_1 (e^-_{\lambda_1} \to e^- _{\lambda_3})$}
  \label{fig:10}
\end{figure}
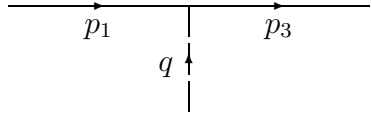

The vertex factor $J_1 (e^\pm _{\lambda_1} \to e^\pm _{\lambda_3} )$
describing  the transition of a $e^\pm$ with  momentum $p_1$ and
helicity $\lambda_1$ to a $e^\pm$ with momentum $p_3$ and
helicity $\lambda_3$  (Fig.~\ref{fig:10})
is equal to
\begin{equation}
J_1 (e^\pm _{\lambda_1} \to e^\pm _{\lambda_3}) =
\sqrt{8\pi \alpha}\, \delta_{\lambda_1 \lambda_3} \, {\mathrm e}^{i
(\lambda_3 \varphi_3- \lambda_1 \varphi_1)} \,.
\label{12}
\end{equation}
This vertex contributes to the transition amplitudes of
Figs.~\ref{fig:1}, \ref{fig:3}, \ref{fig:5}, \ref{fig:6} and \ref{fig:8}.
Its derivation is presented in detail in Sec.~\ref{sec:3.2}.
The azimuthal angle of the initial state lepton $\varphi_1$
can be chosen equal to zero.

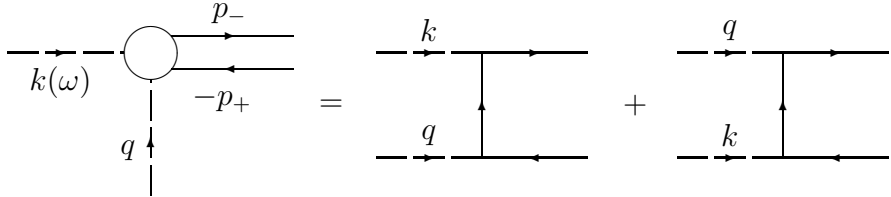
\begin{figure}[!htb]
  \centering
\unitlength=2.00mm
\begin{picture}(61.00,16.00)(2.50,2.50)
\put(12.00,12.00){\circle{3.40}}
\put( 2.50,12.00){\line(1,0){2.0}}
\put( 5.00,12.00){\line(1,0){2.0}}
\put( 5.00,12.00){\vector(1,0){1.40}}
\put( 7.50,12.00){\line(1,0){2.0}}
\put(10.00,12.00){\line(1,0){0.2}}
\put(12.00, 2.50){\line(0,1){2.0}}
\put(12.00, 5.00){\line(0,1){2.0}}
\put(12.00, 5.00){\vector(0,1){1.40}}
\put(12.00, 7.50){\line(0,1){2.0}}
\put(12.00,10.00){\line(0,1){0.2}}
\put(13.40,10.90){\line(1,0){ 8.1}}
\put(13.40,13.10){\vector(1,0) {4.25}}
\put(13.40,13.10){\line(1,0){ 8.10}}
\put(21.50,10.90){\vector(-1,0){4.65}}
\put(17.25,14.60){\makebox(0,0)[cc]{$p_-$}}
\put(16.75, 8.90){\makebox(0,0)[cc]{$-p_+$}}
\put( 6.00,10.00){\makebox(0,0)[cc]{$k(\omega)$}}
\put(10.50, 5.50){\makebox(0,0)[cc]{$q$}}
\put(24.00, 8.50){\makebox(0,0)[cc]{$=$}}
\put(27.00, 5.00){\line(1,0){2.0}}
\put(29.50, 5.00){\vector(1,0){1.40}}
\put(29.50, 5.00){\line(1,0){2.0}}
\put(32.00, 5.00){\line(1,0){2.0}}
\put(27.00,12.00){\line(1,0){2.0}}
\put(29.50,12.00){\vector(1,0){1.40}}
\put(29.50,12.00){\line(1,0){2.0}}
\put(32.00,12.00){\line(1,0){2.0}}
\put(34.00, 5.00){\line(0,1){ 7.0}}
\put(34.00, 5.00){\vector(0,1){3.90}}
\put(34.00,12.00){\line(1,0){ 7.0}} 
\put(34.00,12.00){\vector(1,0){3.90}}
\put(34.00, 5.00){\line(1,0){ 7.0}} 
\put(41.00, 5.00){\vector(-1,0){3.90}}
\put(30.50,13.50){\makebox(0,0)[cc]{$k$}}
\put(30.50, 6.50){\makebox(0,0)[cc]{$q$}}
\put(44.20, 8.50){\makebox(0,0)[cc]{$+$}}
\put(47.00, 5.00){\line(1,0){2.0}}
\put(49.50, 5.00){\vector(1,0){1.40}}
\put(49.50, 5.00){\line(1,0){2.0}}
\put(52.00, 5.00){\line(1,0){2.0}}
\put(47.00,12.00){\line(1,0){2.0}}
\put(49.50,12.00){\vector(1,0){1.40}}
\put(49.50,12.00){\line(1,0){2.0}}
\put(52.00,12.00){\line(1,0){2.0}}
\put(54.00, 5.00){\line(0,1){ 7.0}}
\put(54.00, 5.00){\vector(0,1){3.90}}
\put(54.00,12.00){\line(1,0){ 7.0}}
\put(54.00,12.00){\vector(1,0){3.90}}
\put(54.00, 5.00){\line(1,0){ 7.0}}
\put(61.00, 5.00){\vector(-1,0){3.90}}
\put(50.50,13.50){\makebox(0,0)[cc]{$q$}}
\put(50.50, 6.50){\makebox(0,0)[cc]{$k$}}

\end{picture}
  \caption{ Amplitude $\gamma \gamma^* \to l^+ l^-$ for
  $J_1 (\gamma_{\lambda} \to e^+_{\lambda_+} e^-_{\lambda_-})$}
  \label{fig:11}
\end{figure}

The vertex factor for the transition of a real photon with energy
$\omega$ and helicity $\lambda$ into a lepton pair
(Fig.~\ref{fig:11},
which corresponds to the upper
block of Figs.~\ref{fig:1}, \ref{fig:2}, \ref{fig:7}
and lower block of Fig.~\ref{fig:2}) is equal to
\begin{eqnarray}
J_1 (\gamma_{\lambda} \to e^+_{\lambda_+} e^-_{\lambda_-})\,& =&
 i 4\pi \alpha\, {\sqrt{x_+ x_-}} \left[ (x_+ - x_-
+2\lambda_+ \lambda ) \sqrt 2 {\bf {Q e}}_\lambda \,
\delta_{\lambda_+ , - \lambda_-} \,- \right. \nonumber \\
&  & \left.- 2m R\delta_{\lambda_+, \lambda_-}\, \delta_{\lambda ,
2\lambda_+}\right] \, {\mathrm e}^{i
(\lambda_+ \varphi_+ + \lambda_- \varphi_-)} \,,
\label{13}
\end{eqnarray}
where $x_\pm = E_\pm / \omega$ are the lepton energy fractions. The
quantities ${\bf Q}$  and $R$ are defined in  (\ref{7}) with
\begin{equation}
{\bf u} = {\bf p}_{+ \perp}  \,, \;\;
{\bf v} = {\bf p}_{- \perp}  \,, \;\;
a= m^2 +{\bf u}^2, \;\; b= m^2 +{\bf v}^2 .
\label{14}
\end{equation}

\begin{figure}[!htb]
  \centering
\unitlength=2.00mm
\begin{picture}(61.00,16.00)(2.50,2.50)
\put(12.00,12.00){\circle{3.40}}
\put( 2.50,10.90){\line(1,0){8.10}}
\put( 2.50,10.90){\vector(1,0){5.15}}
\put(12.00, 2.50){\line(0,1){2.0}}
\put(12.00, 5.00){\line(0,1){2.0}}
\put(12.00, 5.00){\vector(0,1){1.40}}
\put(12.00, 7.50){\line(0,1){2.0}}
\put(12.00,10.00){\line(0,1){0.2}}
\put(13.40,10.90){\line(1,0){ 8.1}}
\put(13.40,10.90){\vector(1,0) {5.00}}
\put(13.40,13.10){\line(1,0){0.6}}
\put(14.50,13.10){\line(1,0){2.0}}
\put(17.00,13.10){\line(1,0){2.0}}
\put(19.50,13.10){\line(1,0){2.0}}
\put(17.00,13.10){\vector(1,0){1.40}}
\put(17.50,14.60){\makebox(0,0)[cc]{$k$}}
\put(17.50, 8.90){\makebox(0,0)[cc]{$p_3$}}
\put( 7.00, 8.90){\makebox(0,0)[cc]{$p_1$}}
\put(10.50, 5.50){\makebox(0,0)[cc]{$q$}}
\put(24.00, 8.50){\makebox(0,0)[cc]{$=$}}
\put(27.00, 8.50){\line(1,0){7.0}}
\put(27.00, 8.50){\vector(1,0){2.90}}
\put(31.66, 8.50){\vector(1,0){2.90}}
\put(34.00, 8.50){\line(1,0){7.0}}
\put(34.00, 8.50){\vector(1,0){5.00}}
\put(31.66, 2.50){\line(0,1){2.00}}
\put(31.66, 5.00){\line(0,1){2.00}}
\put(31.66, 7.50){\line(0,1){1.00}}
\put(31.66, 5.00){\vector(0,1){1.40}}
\put(36.34,12.50){\line(0,1){2.00}}
\put(36.34,10.00){\line(0,1){2.00}}
\put(36.34, 8.50){\line(0,1){1.00}}
\put(36.34,10.00){\vector(0,1){1.40}}
\put(34.84,11.20){\makebox(0,0)[cc]{$k$}}
\put(30.16, 5.50){\makebox(0,0)[cc]{$q$}}
\put(44.20, 8.50){\makebox(0,0)[cc]{$+$}}
\put(47.00, 8.50){\line(1,0){ 7.0}}
\put(47.00, 8.50){\vector(1,0){2.90}}
\put(51.66, 8.50){\vector(1,0){2.90}}
\put(54.00, 8.50){\line(1,0){ 7.0}}
\put(54.00, 8.50){\vector(1,0){5.00}}
\put(50.16,11.20){\makebox(0,0)[cc]{$k$}}
\put(54.84, 5.50){\makebox(0,0)[cc]{$q$}}
\put(56.34, 2.50){\line(0,1){2.00}}
\put(56.34, 5.00){\line(0,1){2.00}}
\put(56.34, 7.50){\line(0,1){1.00}}
\put(56.34, 5.00){\vector(0,1){1.40}}
\put(51.66,12.50){\line(0,1){2.00}}
\put(51.66,10.00){\line(0,1){2.00}}
\put(51.66, 8.50){\line(0,1){1.00}}
\put(51.66,10.00){\vector(0,1){1.40}}
\end{picture}
  \caption{ Amplitude $e^- \gamma^* \to  e^- \gamma$ for
  $J_1 (e^- _{\lambda_1} \to e^- _{\lambda_3}\,\gamma_\lambda)$}
  \label{fig:12}
\end{figure}
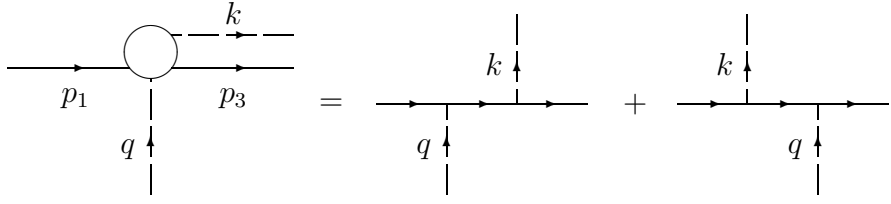

The vertex factor for the transition of an electron or positron
$e^\pm$ to $e^\pm$  and a photon with momentum (energy) $k
(\omega)$ and
helicity $\lambda$ (Fig.~\ref{fig:12},
 appearing in  the
upper block of Figs.~\ref{fig:3}, \ref{fig:4} and lower block of
Fig.~\ref{fig:7}) is given by
\begin{eqnarray}
J_1 (e^\pm _{\lambda_1} \to e^\pm _{\lambda_3}\,
\gamma_\lambda)& =&
\pm 4\pi \alpha\, \sqrt{1-x} \left[ \left(\frac{2-x}{x} +2\lambda_1
\lambda \right)\sqrt{2}  {\bf {Q e}}_{-\lambda}
\delta_{\lambda_1 \lambda_3} \,- \right. \nonumber \\
& & - \left. 2 m R\delta_{\lambda_1,- \lambda_3}\,
\delta_{\lambda, 2\lambda_1} \right] \, {\mathrm e}^{i(\lambda_3
\varphi_3-\lambda_1 \varphi_1)} \, ,
\label{15}
\end{eqnarray}
where $x=\omega / E_1$.
The quantities ${\bf Q}$ and $R$ are  of
form (\ref{7}), however
\begin{equation}
{\bf u} = {{\bf k}_\perp \over x} \,, \;\;
{\bf v} = {\bf q}_\perp - {{\bf k}_\perp \over x}, \;\;
a= m^2 +{\bf u}^2, \;\; b= m^2 +{\bf v}^2 .
\label{16}
\end{equation}
Again the azimuthal angle $\varphi_1$ can be set equal to zero.
Note the useful relation for quantities ${\bf Q}$ and $R$ valid
for relations
(\ref{14}) and (\ref{16})
\begin{equation}
{\bf Q}^2 + m^2 R^2 = {{\bf q}_\perp^2 \over ab}\,.
\label{17}
\end{equation}

The amplitude of Fig.~\ref{fig:12} is the cross amplitude to that of
Fig.~\ref{fig:11}. Therefore, the corresponding vertex factors are
connected by the relation
\begin{equation}
J_1 (\gamma \to e^+ e^-) \to - \frac{ J_1(e \to e \gamma)}{x}
\label{18}
\end{equation}
using the  substitution rules
\begin{equation}
{\bf p}_{+\perp} \to \frac{{\bf k}_\perp}{x} \ , \ \
{\bf p}_{-\perp} \to {\bf q}_\perp- \frac{{\bf k}_\perp}{x} \ ,  \ \
x_+ \to \frac{1}{x} \ , \ \ x_- \to - \frac{1-x}{x}\, ,
\label{19}
\end{equation}
\begin{equation}
 \lambda \to - \lambda \ , \ \
\lambda_+ \to - \lambda_1 \ , \ \
\lambda_- \to  \lambda_3 \ , \ \
\varphi_{+,-} \to \varphi_{1,3} \ .
\label{20}
\end{equation}
Here relations~(\ref{19}) are known substitution rules for the unpolarised
cross
sections \cite{lipatov}.
Taking into account the polarisation of the particles,
relations~(\ref{20}) describe additional
substitution rules to be added to (\ref{19}).

Both vertex factors have  a symmetry related to the obvious symmetry
of Fig.~\ref{fig:11} under lepton exchange $l^+ \leftrightarrow
l^-$.  The vertex factor $J_1(\gamma \to e^+ e^-)$ (\ref{13}) changes 
its sign under
the replacements $+ \leftrightarrow -$. Analogously, the vertex
factor $J_1(e \to e \gamma)/\sqrt{1-x}$ (\ref{15}) changes its
sign under
\begin{equation}
{\bf u} \leftrightarrow {\bf v}, \;\;
x \leftrightarrow - {x\over 1-x}\,, \;\;
\lambda_1 \leftrightarrow - \lambda_3\,, \;\;
\varphi_1 \leftrightarrow \varphi_3 \,.
\label{21}
\end{equation}

Eqs.~(\ref{11})-(\ref{16}) completely describe all helicity
amplitudes of Figs.~\ref{fig:1}-\ref{fig:4} and \ref{fig:7}.
They are not only very compact expressions but they are
also convenient for numerical calculations. The reason is that
in their form large
compensating terms are  already cancelled. Indeed, looking for
the behaviour of ${\bf Q}$ and $ R$ in the limit of vanishing
${\bf q}_\perp$ one  immediately finds that (compare with
Eq.~(\ref{17}))
\begin{equation}
| {\bf Q} |, \  R \propto
 |{\bf q}_\perp| \ \ {\mathrm at} \ \ |{\bf q}_\perp| \to 0 \ .
\label{22}
\end{equation}
This completes the summary of previous results.

\section{Two photon mechanism for the lepton pair production}
\label{sec:3}
\subsection{Sudakov variables}
\label{sec:3.1}

Let us consider  the block diagram of Fig.~\ref{fig:5}. The
4--momenta (energies) of the final electron and positron
are denoted by
$p_3(E_3)$ and $p_4(E_4)$, respectively, those of the produced
muons $\mu^\mp$ by
$p_\mp(E_\mp)$.
The azimuthal angles of the final
particles are $\varphi_{3,4,\pm}$, the polar angles of the final
electron and muons with respect to the $z$-axis  are
$\theta_{3,\pm}$, the polar angles of the final positron with
respect to the ($-z$)-axis is $\theta_{4}$. The electron  and
the muon mass are denoted by $m$ and $M$, respectively.

There are three different kinematic regions to be distinguished:
\vspace{-5mm}
\begin{enumerate}
\item
Electron fragmentation region \\
The particles in the
produced pair move along the initial electron direction (inside
the first jet) with energy $\sim E_1$, and the scattered positron
loses a small fraction of its energy. If we introduce the energy
fractions of the final particles as
\begin{equation}
x_{3,\pm}= \frac{E_{3,\pm}}{E_1} \ , \ \
x= x_+ +x_- \ ,
\label{23}
\end{equation}
these quantities are of the order of 1.
\item
Positron fragmentation region \\
 It is obtained from the electron fragmentation
region  by substituting $e^- \leftrightarrow e^+$.
\item
Region of soft particle production \\
In that region $x_{\pm} \ll 1$,
this case will be discussed in Sec.~\ref{sec:5}.
\end{enumerate}
\vspace{-5mm}
Throughout the paper we treat the electron fragmentation region in detail.

Let us introduce the 4-momenta $k=p_1-p_3$
of the virtual photon $\gamma_k^*$ inside block $J_1$
and $q =p_2-p_4$ of
the virtual photon
$\gamma^*$ connecting  the blocks $J_1$ and $J_2$. We
decompose the 4--vectors $p_i$ ($i=1-4,\pm$), $k$ and $q$ into
components in the plane of the 4--vectors $p$ and $p'$ (see
(\ref{6})) and in the plane orthogonal to them
\begin{eqnarray}
p_i &=& \alpha_i p' + \beta_i p + p_{i\perp} \; , \nonumber \\
k=p_1-p_3 &=& \alpha_k p' +\beta_k p + k_\perp \, , \\
q=p_2-p_4 &=& \alpha_q p' +\beta_q p + q_\perp \, . \nonumber
\label{24}
\end{eqnarray}
The parameters $\alpha$ and $\beta$ are the so called Sudakov
parameters. In the used reference frame the 4--vectors
$p_{i\perp}$, $k_\perp$ and $q_\perp$ have $x$ and $y$ components
only, e.g.
\begin{equation}
q_\perp = (0,\, q_x ,\, q_y, \,0) =
(0,\, {\bf q}_\perp ,\, 0) \,, \;\; \; q^2_\perp =- {\bf q}^2_\perp\, .
\label{25}
\end{equation}

In this jet--like kinematics we note the useful relation
\begin{equation}
p_j^2 = m_j^2 = s\alpha_j \beta_j -{\bf p}^2 _{j\perp} \; .
\label{26}
\end{equation}
for the final state particles (in our case $j=3,4,\pm$) which is valid in
general.
The 4--vectors $p_l$
of particles from the
first jet (here $l=3,\pm$)  obtain  large components along $p_1$ and  small
ones along $p_2$.
Therefore, in the limit  $s\to \infty$ (with accuracy
(\ref{3})) the quantities $\beta_l =2p_l p' / s = E_l / E_1 $ are
finite whereas $ \alpha_l =2p_l p / s = (m^2_l +{\bf p}_{l\perp}^2)
 / (s \beta_l)$ are small\footnote
 { Analogously, for a
4--vector $p_l$ (here $l=4$) of a particle from the second jet the quantity
$\alpha_l = E_l / E_2 $ is finite, $ \beta_l  = (m^2_l +{\bf
p}_{l \perp}^2)/(s\alpha_l ) $ is small.}.

In particular, for our case of the
electron fragmentation region (block diagram of Fig.~\ref{fig:5}, $l=3,\pm$)
the Sudakov parameters
\begin{equation}
\beta_1 = 1,\; \beta_l= x_l,\;
\alpha_2 =1,\; \alpha_4 =\alpha_2-\alpha_q \approx 1,\;
\beta_k = \beta_1 -\beta_3 = 1-x_3 \approx x
\label{27}
\end{equation}
are finite at $s\to \infty$, whereas the parameters
$$
\alpha_1 ={m^2\over s},\;
\alpha_l  = {m^2_l +{\bf p}_{l \perp}^2 \over s x_l} , \;
\beta_2 ={m^2\over s},\;
\beta_4  = {m^2 +{\bf p}_{4 \perp}^2 \over s} ,
$$
\begin{equation}
\alpha_k =\alpha_1 - \alpha_3,\;
\alpha_q = \alpha_+ + \alpha_- +\alpha_3 - \alpha_1,\;
\beta_q = \beta_2-\beta_4
\label{28}
\end{equation}
are small. Therefore, the energy of the final and initial positrons
are approximately equal $E_4 \approx E_2$.
The energy fractions $x$ and $x_l$ are defined in Eq.~(\ref{23}).

Next, we derive a useful expression for the virtualities $k^2$ and
$q^2$. From the obvious relations $(p_1 -k)^2 =p_3^2 $ and
$(p_2-q)^2 = p^2_4$ or $2p_1 k = s\alpha_k + m^2 \beta_k =k^2=
s\alpha_k \beta_k - {\bf k}^2_\perp$ and $2p_2 q = s\beta_q + m^2
\alpha_q =q^2= s\alpha_q \beta_q - {\bf q}^2_\perp$ we obtain
\begin{equation}
k^2 = - {{\bf k}^2_\perp +m^2 \beta^2_k \over
1-\beta_k}\,, \;\;
q^2 = - {{\bf q}^2_\perp +m^2 \alpha^2_q \over
1-\alpha_q}\,.
\label{29}
\end{equation}
According to Eqs. (\ref{11}) the amplitude $e^-e^+ \to e^-e^+
\mu^- \mu^+$ for this kinematics has the form (\ref{4}) where the
vertex factors $J_1$ and $J_2$ have to be determined.

\subsection{Calculation of $J_2$}
\label{sec:3.2}
To clarify some points in the further calculations we start with a
detailed derivation
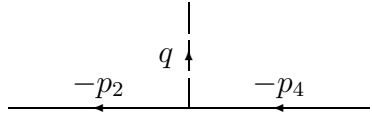
\begin{figure}[!htb]
  \centering
\unitlength=2.00mm
\begin{picture}(24.00,7.00)
\put( 0.00, 0.00){\line(1,0){12.00}}
\put(12.00, 0.00){\vector(-1,0){6.4}} 
\put(12.00, 0.00){\line(0,1){2.0}}
\put(12.00, 2.50){\line(0,1){2.0}}
\put(12.00, 5.00){\line(0,1){2.0}}
\put(12.00, 2.50){\vector(0,1){1.40}}
\put(24.00, 0.00){\vector(-1,0){6.4}} 
\put(12.00, 0.00){\line(1,0){12.0}}
\put(6.00,1.50){\makebox(0,0)[cc]{$-p_2$}}
\put(18.00,1.50){\makebox(0,0)[cc]{$-p_4$}}
\put(10.50,3.30){\makebox(0,0)[cc]{$q$}}
\end{picture}
  \caption{Amplitude $e^+ \gamma^* \to  e^+$ for
  $J_2(e_{\lambda_2}^+ \to e_{\lambda_4}^+)$}
  \label{fig:13}
\end{figure}
of the vertex factor $J_2$ (Fig.~\ref{fig:13}) which is equal to
\begin{equation}
J_2(e_{\lambda_2}^+ \to e_{\lambda_4}^+)
={\sqrt{8\pi\alpha}\over s}\bar v_2\, \hat p\, v_4 \,  .
\label{30}
\end{equation}
The spinor $v_j$  ($j= 2,4$) corresponds to a
positron with 4-momentum $p_j$ and helicity $\lambda_j$. Using
the explicit formulae for  spinors (see Appendix A) we have
$$
J_2 = {\sqrt{8\pi \alpha}\over s} \, E_1 \left( \sqrt{E_2+m}
+\sqrt{E_2-m}\right) \left[\left(\sqrt{E_4+m}
+\sqrt{E_4-m}\right) \cos{{\theta_4\over 2}}\, {\mathrm
e}^{-i\lambda_4 \varphi_4} \; \delta_{\lambda_2 \lambda_4} +
\right.
$$
\begin{equation}
\left.+2\lambda_2
\left(\sqrt{E_4+m} -\sqrt{E_4-m}\right) \sin{{\theta_4\over 2}}\,
{\mathrm e}^{i\lambda_4 \varphi_4} \; \delta_{\lambda_2,\,-
\lambda_4} \right] \,.
\label{31}
\end{equation}
This expression is simplified by omitting terms of the order  $O(m^2 /
E^2_j)$
while keeping terms of the order  $O(m / E_j)$
\begin{equation}
J_2 = \sqrt{8\pi \alpha}
\left[ \cos{{\theta_4\over 2}}
{\mathrm e}^{-i\lambda_4 \varphi_4} \; \delta_{\lambda_2 \lambda_4} +
\lambda_2
{m\over E_4} \,\sin{{\theta_4\over 2}}\,
{\mathrm e}^{i\lambda_4 \varphi_4} \; \delta_{\lambda_2,\,- \lambda_4} \right]
\,.
\label{32}
\end{equation}

Note that in the considered region of small angles (\ref{2}), the term
$\sim m / E_4$ has an additional smallness $\sim \theta_4$. Therefore,
omitting terms of the order $O(m \theta_4 /E_4, \theta_4^2 / E_4^2)$
(Eq.~(\ref{3})) we finally obtain
for the vertex factor of the lower block
\begin{equation}
J_2 = \sqrt{8\pi\alpha} \; {\mathrm e}^{-i\lambda_4\varphi_4} \;
\delta_{\lambda_2\lambda_4}
\label{33}
\end{equation}
from which it follows that the positron conserves its
helicity. The origin of the different sign in the exponent compared to
Eq.~(\ref{12}) is the positron moving opposite to the
direction of the $z$-axis.

\subsection{Calculation of $J_1$}
\label{sec:3.3}
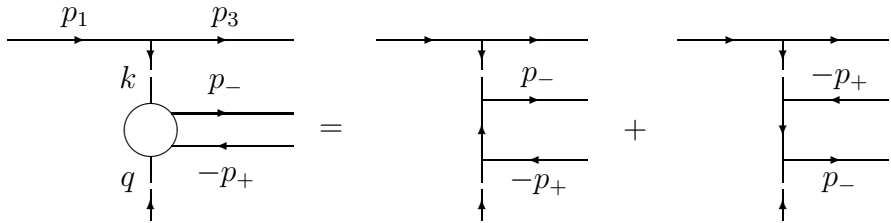
\begin{figure}[!htb]
  \centering
\unitlength=2.00mm
\begin{picture}(61.00,14.00)(2.50,0.00)
\put(12.00, 6.00){\circle{3.40}}
\put(13.40, 4.90){\line(1,0){8.1}}
\put(13.40, 7.10){\vector(1,0) {3.75}}
\put(21.50, 4.90){\vector(-1,0){5.15}}
\put(13.40, 7.10){\line(1,0){8.10}}
\put( 2.50,12.00){\line(1,0){9.50}}
\put( 2.50,12.00){\vector(1,0){5.15}} 
\put(12.00, 0.00){\line(0,1){2.0}}
\put(12.00, 0.00){\vector(0,1){1.40}}
\put(12.00, 2.50){\line(0,1){1.7}}
\put(12.00, 7.80){\line(0,1){1.7}}
\put(12.00,10.00){\line(0,1){2.0}}
\put(12.00,12.00){\vector(0,-1){1.40}}
\put(12.00,12.00){\vector(1,0){5.15}} 
\put(12.00,12.00){\line(1,0){9.5}}
\put( 7.00,13.50){\makebox(0,0)[cc]{$p_1$}}
\put(17.00,13.50){\makebox(0,0)[cc]{$p_3$}}
\put(10.50, 9.50){\makebox(0,0)[cc]{$k$}}
\put(10.50, 2.50){\makebox(0,0)[cc]{$q$}}
\put(17.00, 9.00){\makebox(0,0)[cc]{$p_-$}}
\put(17.00, 3.00){\makebox(0,0)[cc]{$-p_+$}}
\put(24.00, 6.00){\makebox(0,0)[cc]{$=$}}
\put(27.00,12.00){\line(1,0){7.00}}
\put(27.00,12.00){\vector(1,0){3.90}} 
\put(34.00,12.00){\line(1,0){7.00}}
\put(34.00,12.00){\vector(1,0){3.90}} 
\put(34.00, 0.00){\line(0,1){2.0}}
\put(34.00, 0.00){\vector(0,1){1.40}}
\put(34.00, 2.50){\line(0,1){7.0}}
\put(34.00, 4.00){\vector(0,1){2.40}}
\put(34.00,10.00){\line(0,1){2.0}}
\put(34.00,12.00){\vector(0,-1){1.40}}
\put(34.00, 8.00){\line(1,0){7.00}}
\put(34.00, 8.00){\vector(1,0){3.90}} 
\put(34.00, 4.00){\line(1,0){7.00}}
\put(41.00, 4.00){\vector(-1,0){3.90}} 
\put(37.80, 9.50){\makebox(0,0)[cc]{$p_-$}}
\put(37.80, 2.50){\makebox(0,0)[cc]{$-p_+$}}
\put(44.20, 6.00){\makebox(0,0)[cc]{$+$}}
\put(47.00,12.00){\line(1,0){7.00}}
\put(47.00,12.00){\vector(1,0){3.90}} 
\put(54.00,12.00){\line(1,0){7.00}}
\put(54.00,12.00){\vector(1,0){3.90}} 
\put(54.00, 0.00){\line(0,1){2.0}}
\put(54.00, 0.00){\vector(0,1){1.40}}
\put(54.00, 2.50){\line(0,1){7.0}}
\put(54.00, 8.00){\vector(0,-1){2.40}}
\put(54.00,10.00){\line(0,1){2.0}}
\put(54.00,12.00){\vector(0,-1){1.40}}
\put(54.00, 8.00){\line(1,0){7.00}}
\put(61.00, 8.00){\vector(-1,0){3.90}} 
\put(54.00, 4.00){\line(1,0){7.00}}
\put(54.00, 4.00){\vector(1,0){3.90}} 
\put(57.80, 9.50){\makebox(0,0)[cc]{$-p_+$}}
\put(57.80, 2.50){\makebox(0,0)[cc]{$p_-$}}

\end{picture}
  \caption{Two--photon pair production amplitude $e^- \gamma^* \to  e^- \mu^+ \mu^-$ for
   $J_1(e^-_{\lambda_1} \to e^-_{\lambda_3} \mu^+_{\lambda_+}
   \mu^-_{\lambda_-})$}
 \label{fig:14}
\end{figure}

We present the vertex factor $J_1$ (Fig.~\ref{fig:14})
of the two--photon lepton
pair production from Eq.~(\ref{11}) in the form
\begin{equation}
J_1(e^-_{\lambda_1} \to e^-_{\lambda_3} \mu^+_{\lambda_+}
\mu^-_{\lambda_-})
={\sqrt{2}\over s}\,{(4\pi\alpha)^{3/2} \over k^2}\, \bar u_-
\left[\frac{ \hat p'(\hat k-\hat p_+ +M) \hat I}{2 k p_+ -k^2}+
     \frac{ \hat I (\hat p_- - \hat k + M)\hat p'}{2 k p_- -k^2}
\right] v_+ \,,
\label{34}
\end{equation}
where the expression in square brackets corresponds to the
amplitude of the $\gamma^*_k\gamma^*  \to\mu^+\mu^-$ process of
Fig.~\ref{fig:11} with two virtual photons. The 4--vector
\begin{equation}
I_\mu=\bar u_3\gamma_\mu u_1
\label{35}
\end{equation}
is the current of the $e^-\to e^-\gamma^*_k$ transition.
Comparing $J_1$ (Eq.~(\ref{34})) with the corresponding expression
needed in the
usual photoproduction $\gamma e\to\mu^+\mu^- e$ of
Fig.~\ref{fig:1}
\begin{equation}
J_1( \gamma_\lambda \to  \mu^+_{\lambda_+}\mu^-_{\lambda_-})=
\frac{\sqrt{2}}{s} 4 \pi \alpha \ \bar u_-
\left[\frac{ \hat p'(\hat k-\hat p_+ +M) \hat e}{2 k p_+}+
     \frac{ \hat e (\hat p_- - \hat k + M)\hat p'}{2 k p_-}
\right] v_+
\label{36}
\end{equation}
we note two significant differences: \\
{\it (i)} In the process of Fig.~\ref{fig:5} the photon $\gamma_k^*$ with
4--momentum $k$ is
 virtual, $k^2 \not= 0$, while in the usual photoproduction this
photon is real, $k^2=0$;\\
{\it (ii)} In the  photoproduction $J_1$ includes the photon
polarisation vector $e_\mu$ while in the process of
Fig.~\ref{fig:5} the vertex factor $J_1$ includes the quantity
$\sqrt{4 \pi \alpha}  I_\mu/k^2$.\\
In the further calculation we follow the scheme developed in
Refs.~\cite{kuraevnp85}, \cite{kuraevzp86} taking into account
the noticed differences.

In order to calculate the vertex $J_1$ it is convenient
(using gauge invariance) to
replace the current $I_\mu$ by the 4--vector $V_\mu$  which has no
component along $p$ (just as in photoproduction where we can
choose $e_\mu$ in the form $e_\mu = e_{\mu \perp}$)
\begin{equation}
I_\mu \to V_\mu=I_\mu-{\beta_I\over\beta_k}k_\mu=
\alpha_{V}\, p'_\mu+V_{\perp\mu}\,.
\label{37}
\end{equation}
since $\beta_V=2 V p' /s=0$. Consequently, in the final
expression only $V_\perp$ enters.

Using the Sudakov decomposition of Sec.~\ref{sec:3.1}
the denominators in Eq.~(\ref{34}) can be transformed to
\begin{equation}
2k p_\pm-k^2=
{x\over x_\pm}\,\left(M^2+{\bf r}^2_\pm-{x_+ x_-
\over x^2}\, k^2 \right)
\label{38}
\end{equation}
with
\begin{equation}
r^\mu_\pm= p^\mu_{\pm\perp}-{x_\pm\over x}\,k^\mu_{\perp}
\,, \;\;\;
{\bf r}_+ + {\bf r}_- = {\bf q}_\perp \,.
\label{39}
\end{equation}
It is useful to note that the vectors ${\bf r}_+$ and ${\bf r}_-$
(corresponding to the vectors ${\bf u}$ and ${\bf v}$ in Eq. (\ref{43})
below) are the components of the vectors ${\bf p}_+$ and ${\bf p}_-$
transverse to the vector ${\bf k}$ in full analogy with the
corresponding Eq.~(\ref{14}) for the photoproduction of Fig.~\ref{fig:1}.

In the numerator of the first term $N=\bar u_-\hat p'(\hat k-\hat
p_++ M)\hat V v_+$  the matrix $\hat V$ is transposed to the left,
and using the Dirac equation $(\hat p_+ + M)\, v_+ =0$ we obtain
\begin{equation}
N =\bar u_-\hat p'\,[2V(k-p_+) -\hat V \,\hat k] \,v_+ \,.
\label{40}
\end{equation}
Taking into account
$$
2V(k-p_+)=-2V_\perp r_+ - 2 \frac{x_-}{x^2} \, \beta_I \, k^2
$$
and
$$
\hat p'\,\hat V \, \hat k \, v_+ = -
\frac{x}{x_+} \hat p'\,\hat V_\perp (\hat r_+ +
M)\,v_+
$$
we transform $N$ to
\begin{equation}
N={x\over x_+}\bar u_- \,\hat p' \left[- {2x_+ \over x}V_\perp
r_+ + \hat V_\perp(\hat r_+ +M) -2{ x_+ x_- \over x^3}
\,\beta_I\,k^2\,  \right] v_+ \,.
\label{41}
\end{equation}

With similar transformations for the second term we find
the vertex factor $J_1$ in the compact form
$$
J_1=\sqrt{2}\,{(4\pi\alpha)^{3/2}\over s \, k^2}\, \bar u_-\hat
p'\Lambda v_+ \,,
$$
\begin{equation}
\Lambda=2\,{x_+\over x} \,{\bf V}_\perp{\bf Q}+\hat
V_\perp(\hat Q+ M R)
- 2{ x_+ x_- \over x^3} \,\beta_I\,R\, k^2\,
\label{42}
\end{equation}
with
\begin{equation}
{\bf u} ={\bf r}_+\,, \;
{\bf v} ={\bf r}_-\,, \;
a=M^2 + {\bf u}^2-{x_+ x_-\over x^2}\,k^2 \,, \;
b=M^2 + {\bf v}^2-{x_+ x_-\over x^2}\,k^2 \, .
\label{43}
\end{equation}
The 4--vector $Q$ in the considered reference frame has
transverse components only $Q=(0,{\bf Q},0)$ and the quantities
${\bf Q}$ and $R$ are defined in Eq.~(\ref{7}). Note the
relation (see Eq.~(\ref{7a}) and compare with Eq.~(\ref{17}))
\begin{equation}
{\bf Q}^2+\left(M^2-{x_+ x_-\over x^2}\,k^2 \right)\,
 R^2={{\bf q}^2_{\perp}\over ab}\,.
\label{44}
\end{equation}

To prove the independence of $J_1$ on $s$, we use the explicit
formulae for the spinors $u_-$ and $v_+$ and omit terms of the
order of (\ref{3}) (just as it was done in
Eqs.~(\ref{32})-(\ref{35})). Moreover, to get a simple final
expression for the helicity amplitudes, it is useful to introduce
the helicity eigenvectors  ${\bf e}_\lambda$ (\ref{8}) and to
decompose the vector ${\bf V}_\perp $ in the helicity basis ${\bf
e}_\lambda$
\begin{equation}
{\bf V}_\perp=\sum_{\lambda=\pm 1}({\bf V}_\perp{\bf e}^*_\lambda)\,
{\bf e}_\lambda
= - \sum_{\lambda=\pm 1}({\bf V}_\perp{\bf e}_{-\lambda}) \,
 {\bf e}_\lambda \ .
\label{45}
\end{equation}
As a result, we obtain with  accuracy (\ref{3})
$$
J_1=i\,(4\pi\alpha)^{3/2}{\sqrt{x_+ x_-}\over k^2} \,
{\mathrm e}^{i(\lambda_+\varphi_+ +\lambda_-\varphi_-)}\times
$$
$$
\times \left\{ \sum_\lambda ({\bf V}_\perp{\bf e^*_\lambda})
\left[ \left( {x_+- x_-\over x}
+2 \lambda_+ \lambda \right) \sqrt{2} {\bf {Q e}}_\lambda
 \delta_{\lambda_+,-\lambda_-} -2M
R\delta_{\lambda_+\lambda_-}\delta_{\lambda,2\lambda_+}\right]-
\right.
$$
\begin{equation}
\left. -2\sqrt{2}\,{x_+ x_-\over x^3} \,\beta_I\, R\, k^2
\delta_{\lambda_+,-\lambda_-} \right\} \,.
\label{46}
\end{equation}

For vanishing virtuality ($k^2 \to 0$) the expression in
the square brackets of this equation coincides with the
corresponding expression in Eq.~(\ref{13}) by identifying $x=1$.

The result (\ref{46}) can be presented with the same accuracy
in another form using the two-component spinors $w_{\lambda_j}$
in which we have to put the polar scattering angles of muons
equal to zero ($\theta_{\pm}=0$)
$$
J_1=i \sqrt{2}(4\pi\alpha)^{3/2}{\sqrt{x_+ x_-}\over
k^2} \times
$$
\begin{equation}
\times w^+_{\lambda_-} \left\{{x_+- x_-\over x}{\bf
Q}{\bf V}_\perp+ i{\mbox{\boldmath$\sigma$}} [{\bf Q}+ M
 R \, {\bf n}_1,{\bf V}_\perp] -
2\,{x_+ x_-\over x^3}\,\beta_I \,R\, k^2 \right\} w_{-\lambda_+}
\label{47}
\end{equation}
with the Pauli matrices $\mbox{\boldmath$\sigma$}$ and the unit
vector
\begin{equation}
{\bf n}_1 = {{\bf {p}}_1  \over |{\bf {p}}_1|}\,.
\label{48}
\end{equation}

The only quantities which remain  to be calculated are $\beta_I$ and
${\bf V}_\perp{\bf {e}}^*_\lambda$.
Setting the azimuthal angle of the incoming electron equal to zero ($\varphi_1=0$)
we find
\begin{equation}
\beta_I=\frac{2}{s} \, p'_\mu I^\mu={2\over s}\bar u_3\hat p' u_1=
2\sqrt{1- x}\
{\mathrm e}^{i\lambda_3\varphi_3} \ \delta_{\lambda_1\lambda_3}\,,
\label{49}
\end{equation}
and
$$
{\bf V}_\perp{\bf {e}}^*_\lambda
=-\lambda \sqrt{2 E_1 E_3} \,
{\mathrm e}^{i\lambda_3\varphi_3} \times
$$
\begin{equation}
\times \left[ \left( {1-x \over x} +\delta_{\lambda,
2\lambda_1} \right) \, \theta_3 {\mathrm e}^{-i\lambda\varphi_3}
\,\delta_{\lambda_1\lambda_3}+ \lambda\left({m \over
E_3 }-{m \over E_1}\right) \delta_{\lambda,
2\lambda_1} \  \delta_{\lambda_1,-\lambda_3} \right] \,.
\label{50}
\end{equation}
The corresponding expressions suitable for crossing can be obtained
by holding the energy fraction $x_1(=1)$, the polar $\theta_1(=0)$
and azimuthal $\varphi_1(=0)$ angles of the incoming electron as
free parameters.
In that case one obtains ($x_3=1-x$)
\begin{equation}
\beta_I=
2\sqrt{x_1 x_3}\
{\mathrm e}^{i(\lambda_3\varphi_3-\lambda_1\varphi_1)} \
\delta_{\lambda_1\lambda_3} \,,
\label{49a}
\end{equation}
$$
{\bf V}_\perp {\bf {e}}^*_\lambda=-{\bf V}_\perp {\bf {e}}_{-\lambda}
=-\lambda \sqrt{2 E_1 E_3} \,
{\mathrm e}^{i(\lambda_3\varphi_3-\lambda_1\varphi_1)} \times
$$
\begin{equation}
\times \left\{\left[
\left( \frac{x_1}{x} + \delta_{\lambda,-2\lambda_1}\right) \, \theta_1  \,
{\mathrm e}^{-i\lambda\varphi_1}+
\left( \frac{x_3}{x} + \delta_{\lambda, 2\lambda_1} \right) \, \theta_3 \,
{\mathrm e}^{-i\lambda\varphi_3} \right]
\delta_{\lambda_1\lambda_3} -
\right.
\label{50a}
\end{equation}
$$
\left.
- \lambda \left({m \over E_3 }-{m \over E_1}\right)
\delta_{\lambda, 2\lambda_1}\  \delta_{\lambda_1,-\lambda_3} \right\} \,.
$$

The newly calculated vertex factor $J_1$ has again a symmetry
related to the obvious symmetry of Fig.~\ref{fig:14} under lepton
exchange $\mu^+ \leftrightarrow \mu^-$. It changes its sign
when replacing indices $+ \leftrightarrow -$.

\section{Bremsstrahlung mechanism for the lepton pair 
production}
\label{sec:4}
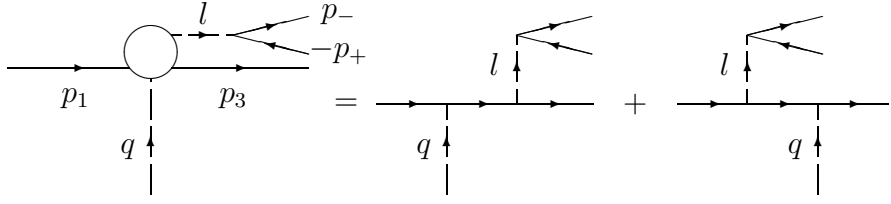
\begin{figure}[!htb]
  \centering
\unitlength=2.00mm
\begin{picture}(61.00,16.00)(2.50,2.50)
\put(12.00,12.00){\circle{3.40}}
\put( 2.50,10.90){\line(1,0){8.10}}
\put( 2.50,10.90){\vector(1,0){5.15}}
\put(12.00, 2.50){\line(0,1){2.0}}
\put(12.00, 5.00){\line(0,1){2.0}}
\put(12.00, 5.00){\vector(0,1){1.40}}
\put(12.00, 7.50){\line(0,1){2.0}}
\put(12.00,10.00){\line(0,1){0.2}}
\put(17.50,13.10){\line(4,1){5.0}}
\put(17.50,13.10){\vector(4,1){2.9}}
\put(17.50,13.10){\line(4,-1){5.0}}
\put(22.50,11.85){\vector(-4, 1){2.9}}
\put(13.40,10.90){\line(1,0){ 9.1}}
\put(13.40,10.90){\vector(1,0) {5.00}}
\put(13.40,13.10){\line(1,0){0.6}}
\put(14.50,13.10){\line(1,0){2.0}}
\put(17.00,13.10){\line(1,0){0.5}}
\put(14.50,13.10){\vector(1,0){1.40}}
\put(15.50,14.60){\makebox(0,0)[cc]{$l$}}
\put(17.50, 8.90){\makebox(0,0)[cc]{$p_3$}}
\put( 7.00, 8.90){\makebox(0,0)[cc]{$p_1$}}
\put(10.50, 5.50){\makebox(0,0)[cc]{$q$}}
\put(25.00, 8.50){\makebox(0,0)[cc]{$=$}}
\put(24.50,14.50){\makebox(0,0)[cc]{$p_-$}}
\put(24.50,12.00){\makebox(0,0)[cc]{$-p_+$}}
\put(27.00, 8.50){\line(1,0){7.0}}
\put(27.00, 8.50){\vector(1,0){2.90}}
\put(31.66, 8.50){\vector(1,0){2.90}}
\put(34.00, 8.50){\line(1,0){7.4}}
\put(34.00, 8.50){\vector(1,0){5.30}}
\put(31.66, 2.50){\line(0,1){2.00}}
\put(31.66, 5.00){\line(0,1){2.00}}
\put(31.66, 7.50){\line(0,1){1.00}}
\put(31.66, 5.00){\vector(0,1){1.40}}
\put(36.34,12.50){\line(0,1){0.60}}
\put(36.34,10.00){\line(0,1){2.00}}
\put(36.34, 8.50){\line(0,1){1.00}}
\put(36.34,10.00){\vector(0,1){1.40}}
\put(36.34,13.10){\line(4,1){5.0}}
\put(36.34,13.10){\vector(4,1){2.9}}
\put(36.34,13.10){\line(4,-1){5.0}}
\put(41.34,11.85){\vector(-4, 1){2.9}}
\put(34.84,11.20){\makebox(0,0)[cc]{$l$}}
\put(30.16, 5.50){\makebox(0,0)[cc]{$q$}}
%
\put(44.20, 8.50){\makebox(0,0)[cc]{$+$}}
\put(47.00, 8.50){\line(1,0){ 7.0}}
\put(47.00, 8.50){\vector(1,0){2.90}}
\put(51.66, 8.50){\vector(1,0){2.90}}
\put(54.00, 8.50){\line(1,0){ 7.0}}
\put(54.00, 8.50){\vector(1,0){5.20}}
\put(50.16,11.20){\makebox(0,0)[cc]{$l$}}
\put(54.84, 5.50){\makebox(0,0)[cc]{$q$}}
%
\put(56.34, 2.50){\line(0,1){2.00}}
\put(56.34, 5.00){\line(0,1){2.00}}
\put(56.34, 7.50){\line(0,1){1.00}}
\put(56.34, 5.00){\vector(0,1){1.40}}
\put(51.66,12.50){\line(0,1){0.60}}
\put(51.66,10.00){\line(0,1){2.00}}
\put(51.66, 8.50){\line(0,1){1.00}}
\put(51.66,10.00){\vector(0,1){1.40}}
\put(51.66,13.10){\line(4,1){5.0}}
\put(51.66,13.10){\vector(4,1){2.9}}
\put(51.66,13.10){\line(4,-1){5.0}}
\put(56.66,11.85){\vector(-4, 1){2.9}}
\end{picture}
  \caption{Bremsstrahlung pair production amplitude
   $e^- \gamma^* \to  e^- \mu^+ \mu^-$ for
   $J_1(e^-_{\lambda_1} \to e^-_{\lambda_3} \mu^+_{\lambda_+}
   \mu^-_{\lambda_-})$}
 \label{fig:15}
\end{figure}

Let us consider the block diagram of Fig.~\ref{fig:6}. The
corresponding amplitude has the form (\ref{11}) where the vertex
factor $J_2$ is the same as in Eq.~(\ref{33}). The vertex factor
$J_1$ (Fig.~\ref{fig:15})
we present in the form
\begin{equation}
J_1(e^-_{\lambda_1} \to e^-_{\lambda_3} \mu^+_{\lambda_+}
   \mu^-_{\lambda_-})
={\sqrt {2}\over s}\,{(4\pi\alpha)^{3/2} \over l^2}\, \bar u_3
\left[\frac{ \hat p'(-\hat l + \hat p_1+m) \hat I}{2 l p_1 -l^2}
+ \frac{ \hat I (\hat p_3 + \hat l + m)\hat p'}{-2l p_3 -l^2}
\right] u_1
\label{51}
\end{equation}
where the expression in square brackets corresponds to the
amplitude of the Compton scattering of the type
of Fig.~\ref{fig:12} with the virtual initial photon $q$ and the
virtual final photon
$$
l = p_+ + p_- = \alpha_l p' +\beta_l p +l_\perp \, .
$$
Now the 4--vector $I_\mu$
denotes the current of the $\gamma^*\to\mu^+\mu^-$
transition
\begin{equation}
I_\mu=\bar u_-\gamma_\mu v_+ \, .
\label{52}
\end{equation}
As in Sec.~\ref{sec:3.3} it is convenient to
replace this current by the 4-vector $V_\mu$ without the
component along $p$ (compare with Eq.~(\ref{37}))
\begin{equation}
I_\mu \; \to \; V_\mu =I_\mu -{\beta_I\over\beta_l} \; l_\mu =
\alpha_V p'_\mu + V_{\perp \mu}
\label{53}
\end{equation}
because $\beta_V = 2V p' /s =0$.

It is easily to see that the vertex factor (\ref{51}) with its current
(\ref{52}) can be obtained from
Eq.~(\ref{34}) and (\ref{35})  substituting
\begin{equation}
p_+ \to - p_1\,, \;\;
p_- \to p_3\,, \;\;
k \to - l\,,\;\;
M \to m\,,\;\;
v_+ \leftrightarrow u_1\,,\;\;
\bar u_- \leftrightarrow \bar u_3 \,.
\label{54}
\end{equation}
This means that the final expression for $J_1$ can be found from
Eq.~(\ref{46}) by the following substitution rules (compare with
Eqs.~(\ref{19})-(\ref{20}))
$$
k^2 \to l^2\,,\;\;
M \to m \,,\;\;
{\bf r}_{+} \to \frac{{\bf l}_\perp}{x} \ , \ \
{\bf r}_{-} \to {\bf q}_\perp- \frac{{\bf l}_\perp}{x} \ ,
$$
\begin{equation}
\beta_+ =x_+ \to -\beta_1 = -1 \ , \ \
\beta_- = x_- \to \beta_3 =x_3 \approx 1-x \,,\;\;
\beta_k \approx x \to -\beta_l = -x\,,
\label{55}
\end{equation}
and
\begin{equation}
 \lambda \to - \lambda \ , \ \
\lambda_+ \to - \lambda_1 \ , \ \
\lambda_- \to  \lambda_3 \ , \ \
\varphi_{+,-} \to \varphi_{1,3} \ .
\label{56}
\end{equation}
Using these rules we find (compare with Eq. (\ref{46}))
$$
J_1=(4\pi\alpha)^{3/2}\ {\sqrt{1-x}\over l^2}
\; {\mathrm e}^{i(\lambda_3\varphi_3 -\lambda_1\varphi_1)} \times
$$
$$
\times \left\{\sum_\lambda({\bf V}_\perp{\bf e}_\lambda) \left[
\left( {2-x \over x} +2\lambda_1\lambda\, \right)\,\sqrt{2}
{\bf {Qe}}^*_\lambda\, \delta_{\lambda_1\lambda_3}+ 2m R\
\delta_{\lambda_1,-\lambda_3}\delta_{\lambda,2\lambda_1} \right]
-\right.
$$
\begin{equation}
\left. -2\sqrt{2}\, {1-x\over x^3}\, \beta_I R\,l^2\,
\delta_{\lambda_1\lambda_3} \right\} \,.
\label{57}
\end{equation}
where
\begin{equation}
{\bf u}={{\bf l}_\perp\over x}, \;
{\bf v}={\bf q}_\perp-{{\bf l}_\perp \over x} ,\;
a=m^2+{\bf u}^2+{1-x \over x^2}l^2 \, ,\;
b=m^2+{\bf v}^2+{1-x\over x^2}l^2
\label{58}
\end{equation}
and the quantities ${\bf Q}$ and $R$ are given in
Eq.~(\ref{7}).

For $l^2 \to 0$ the expression in square brackets in
Eq.~(\ref{57}) coincides with the corresponding expression in
Eq.~(\ref{15}).

Let us point out that
\begin{equation}
l^2 =(p_+ +p_-)^2 = s (\alpha_+ +\alpha_-)\beta_l -
({\bf p}_{+\perp}+{\bf p}_{-\perp}) ^2 =
{(x_+ {\bf p}_{-\perp} -x_-{\bf p}_{+\perp})^2+ M^2 x^2 \over x_+
x_-}\,.
\label{59}
\end{equation}
Note also the useful relation (compare with Eqs.~(\ref{17}) and
(\ref{44}))
\begin{equation}
{\bf Q}^2+\left(m^2+{1-x\over x^2}\,l^2 \right)\,
 R^2={{\bf q}^2_{\perp}\over ab}\,.
\label{60}
\end{equation}

The result (\ref{57}) can also be presented with the same accuracy
in another form using two-component spinors $w_{\lambda_j}$ in
which we have to set the polar scattering angle $\theta_3$ equal
to zero, $\theta_{3}=0$ (compare with Eq.~(\ref{46}))
$$
J_1=\sqrt{2}\,(4\pi\alpha)^{3/2} {\sqrt{1-x}\over l^2 } \,\times
$$
\begin{equation}
w^+_{\lambda_3}\,\left\{{2- x\over x}\, {\bf Q}{\bf V}_\perp+
i {\mbox{\boldmath$\sigma$}} [{\bf Q}  +  m R \,
{\bf n}_1,{\bf V}_\perp]
-2 {1-x\over x^3} \,\beta_I \, R\, l^2 \, \right\}\;
w_{\lambda_1}\, .
\label{61}
\end{equation}

The explicitly calculated quantities $\beta_I$ and
${\bf V}_\perp{\bf e}_\lambda$ are
\begin{equation}
\beta_I=\frac{2}{s} \,p'_\mu I^\mu= \frac{2}{s} \bar u_-\hat p' v_+ =
2i \sqrt{x_+ x_-}\
{\mathrm e}^{i(\lambda_+\varphi_+ + \lambda_-\varphi_-) } \
\delta_{\lambda_+, -\lambda_-}\,,
\label{62}
\end{equation}
and
$$
{\bf V}_\perp {\bf e}_\lambda
=i\lambda \sqrt{2 E_+ E_-} \,
{\mathrm e}^{i(\lambda_+ \varphi_+ + \lambda_- \varphi_-)} \times
$$
\begin{equation}
\times \left\{ \left[
\left( {x_+\over x} -\delta_{\lambda,
2\lambda_-} \right) \, \theta_+ {\mathrm e}^{i\lambda\varphi_+} +
\left( {x_-\over x} -\delta_{\lambda,
2\lambda_+} \right) \, \theta_- {\mathrm e}^{i\lambda\varphi_-}
\right]
\,\delta_{\lambda_+, -\lambda_-}
\right. +
\label{63}
\end{equation}
$$
\left.
\lambda\left({M \over
E_+ }+{M \over E_-}\right) \delta_{\lambda,
2\lambda_+}\, \delta_{\lambda_+,\lambda_-} \right\}
$$
Eqs.~(\ref{49a})-(\ref{50a}) or (\ref{49})-(\ref{50}) can be obtained from
Eqs.~(\ref{62})-(\ref{63}) using the rules
(\ref{55})-(\ref{56}).
\section{Discussion}
\label{sec:5}
The main results of our paper are summarised in Eqs.~(\ref{46}) and
(\ref{57}) (combined with (\ref{4}) and (\ref{33}))
which give the analytical expressions
of all 64 helicity amplitudes
for small angle lepton pair production
in $e^-e^+$ or $\mu^-\mu^+$ collisions.
Since various distributions of this pair production are well-known
(see, for example, reviews \cite{bgms} and \cite{bfkk}), we
briefly discuss only some qualitative features of the obtained
results.

1. As for the processes of Figs.~\ref{fig:1}-\ref{fig:4}, \ref{fig:7},
the obtained
formulae are very compact expressions.
Additionally, they are
convenient for numerical calculations since in their form large
compensating terms are already cancelled. Indeed, the quantities
${\bf Q}$ and $R$, which are defined in Eqs.~(\ref{7}) (with  (\ref{43}) or
(\ref{58})), and the vertex factors   $J_1$ (\ref{46}), (\ref{57})
themselves vanish at
small transverse momentum of the t--channel exchange photon $|{\bf q}_\perp|$
\begin{equation}
| {\bf Q} |, \  R, \; J_1\; \propto
 |{\bf q}_\perp| \ \ {\mathrm at} \ \ |{\bf q}_\perp| \to 0 \ .
\label{64}
\end{equation}
Therefore, the amplitude of the process behaves as
((\ref{4}),(\ref{64}),(\ref{33}),(\ref{29}))
\begin{equation}
M_{fi} \; \propto {|{\bf q}_\perp|\over {\bf q}_\perp^2 +m^2
\alpha_q^2 } \ \ {\mathrm at} \ \ |{\bf q}_\perp| \to 0 \ .
\label{65}
\end{equation}
Here the mass $m$ denotes that of the colliding lepton.

2. The behaviour of $J_1$ for two-photon pair production
(\ref{46}) at small $|{\bf k}_\perp|$ is mainly determined  by the
factor ${\bf V}_\perp /k^2$. Averaging over the spin states of
the initial electron and summing up over the final electron spin
 we find
\begin{equation}
{1\over 2} \sum_{\lambda_1 , \lambda_3}\, V_i V^*_j =
{4\over x^2} k_i k_j - k^2 \, \delta_{ij}; \;\; i,j = x,y
\label{66}
\end{equation}
from which it follows that
\begin{equation}
| J_1 |^2 \propto {(2-2x+x^2) {\bf {k}}^2_\perp +m^2x^4 \over
({\bf k}_\perp^2 +m^2x^2)^2} \ \ {\mathrm at} \ \
|{\bf k}_\perp| \to 0 \ .
\label{67}
\end{equation}

3. Let us consider the amplitudes for bremsstrahlung pair
production which violate electron helicity conservation. They
are proportional to (see Eq.~(\ref{57}))
\begin{equation}
m\,R\, \delta_{\lambda_1, -\lambda_3}\, \delta_{\lambda,
 2\lambda_1} \, .
\label{68}
\end{equation}
Therefore, in such amplitudes the photon helicity $\lambda$ is
strictly connected with the helicity of the initial electron
$\lambda = 2\lambda_1 $. The relative magnitude of these
amplitudes can be estimated as
\begin{equation}
{| J_1 (\lambda_3 = -\lambda_1) |  \over
 | J_1 (\lambda_3 = +\lambda_1) | } \sim
{xm |R| \over | {\bf Q}|} \,.
\label{69}
\end{equation}
This ratio is small at $m/| {\bf p}_{i \perp}|, \; m/ |{\bf
 q}_\perp| \ll 1$,  i.e. in the region of not very small
scattering angles.
The same behaviour show the amplitudes which violate   muon
helicity conservation being proportional to $M\,
\delta_{\lambda_+, \lambda_-}\, \delta_{\lambda, 2\lambda_+}$.

It is interesting to note that the amplitudes which violate
the muon (electron) helicity conservation have a
specific dependence on the
azimuthal angles. This is true both for the two-photon and
bremsstrahlung mechanism:  if $\lambda_+=\lambda_- $
(or $\lambda_1=-\lambda_3)$ then $M_{fi} \sim \exp (i \lambda_+
(\varphi_-+\varphi_+))$ (or $\ \ M_{fi} \sim
\exp (i\lambda_1(\varphi_1+\varphi_3))) $.

4. The two-photon lepton pair production mechanism dominates in the region of
soft particle production mentioned in \ref{sec:3.1}.
The produced leptons have  low energies   compared with
the initial energy $E_1$, i.e. at $x_\pm \ll 1$. In this region
\begin{equation}
{\bf V}_\perp \approx - {\beta_I \over x}\, {\bf k}_\perp
\label{70}
\end{equation}
and the vertex factor $J_1$ (\ref{47}) is considerably simplified
$$
J_1=-i \sqrt{2}(4\pi\alpha)^{3/2}\,{\sqrt{x_+ x_-}\over x}\,
{\beta_I \over k^2} \,
w^+_{\lambda_-}\, \left(A + i{\mbox{\boldmath$\sigma$}} {\bf B}
\right)\, w_{-\lambda_+}\,,
$$
\begin{equation}
A= {x_+- x_-\over x}\,{\bf Q}{\bf k}_\perp -
2\,{x_+ x_-\over x^2}\,R\, {\bf k}^2_\perp \,, \;\;
{\bf B} = [{\bf Q}+ M R \, {\bf n}_1,{\bf k}_\perp]
\,.
\label{71}
\end{equation}
The differential cross section
for unpolarised particles is obtained as follows ($i=1-4,\pm$)
\begin{equation}
d\sigma = {1\over 4} \sum_{\lambda_i} \left|{s\over q^2 } J_1
J_2\right|^2 \, {d\Gamma \over 2s}= 32 {(4\pi \alpha)^4 \over
(q^2 k^2)^2} {x_+ x_- \over x^2}\, \left( A^2 + {\bf B}^2 \right)
\, {s\over 2 }\, d\Gamma \,,
\label{72}
\end{equation}
where $d \Gamma$ is the phase space of the final particles.
Using the relation (\ref{44}), the expression $A^2 + {\bf B}^2$
can be transformed to the symmetric form
$$
A^2 +{\bf B}^2 = {{\bf k}^2_\perp {\bf q}^2_\perp \over ab } -
{x_+ x_- \over x^2 a^2 b^2} \left\{
({\bf k}_\perp {\bf q}_\perp)
({\bf k}_\perp {\bf q}_\perp -
{\bf p}_{+ \perp}^2 - {\bf p}_{- \perp}^2  -2M^2 ) +
({\bf k}_\perp {\bf \Delta })
({\bf q}_\perp {\bf \Delta })
\right\}^2   \,,
$$
\begin{equation}
{\bf \Delta } = {\bf p}_{+ \perp} - {\bf p}_{- \perp}
\label{73}
\end{equation}
which coincides with Eq.~(4.21) in Ref.~\cite{bfkk}.

\vspace{10mm}

{\small

{\it Acknowledgements}. We are grateful to E.~Boos, G.~Kotkin and
Y.~Kurihara for useful discussions. This work is supported
in part by Volkswagen Stiftung (Az. No.  I/72 302) and by Russian
Foundation for Basic Research (code 96-02-19114).
}
\appendix
\section{Definitions of polarised leptonic spinors}
\label{app:A}
A polarisation state of an electron with  momentum $\bf p$,
energy $E=\sqrt{{\bf p}^2+m^2}$ and helicity $\lambda =\pm 1/2$
is described by the bispinor
$$
u^{(\lambda)}_{\bf p}=
\left(
\begin{array}{c}
 \sqrt{E+m}\,w^{(\lambda)}_{\bf n}\\
 \sqrt{E-m}\,(\mbox{\boldmath $ \sigma$} {\bf n})\, w^{(\lambda)}_{\bf n}
\end{array} \right)\,, \;\qquad
{\bf n}={{\bf p}\over \mid{\bf
p}\mid}=(\sin\theta\cos\varphi, \sin\theta\sin\varphi,\cos\theta),
$$
where {$ \boldmath \sigma$} are the Pauli matrices and the
two-component spinors $w^{(\lambda )}_{\bf n}$ obey the
equations
$$
(\mbox{\boldmath $ \sigma$}{\bf n })\,w^{(\lambda )}_{\bf
n}=2\lambda\, w^{(\lambda )} _{\bf n}\,,\;\; w_{\bf
n}^{(\lambda )+}w_{\bf n}^{(\lambda ')}=\delta
_{\lambda\lambda'}\,,
$$
$$
w_{\bf n}^{(1/2)}=\left(
\begin{array}{c}
{\mathrm e}^{-i\varphi /2}\cos{\theta \over 2}  \\
{\mathrm e}^{i \varphi /2}\sin{\theta \over 2}
\end{array} \right) \,, \;\;\;\;
w_{\bf n}^{(-1/2)}=\left(
\begin{array}{c}
 -{\mathrm e}^{-i\varphi /2}\sin{\theta \over 2} \\
{\mathrm e}^{i\varphi /2}\cos{\theta\over 2}
\end{array} \right) \,.
$$
The normalisation conditions are
$$
\bar u^{(\lambda )}_{\bf p} u^{(\lambda ')}_{\bf p}=2m\,\delta
_{\lambda\lambda '}\,,\qquad
\sum _\lambda u^{(\lambda )}_{\bf p} \bar u ^{(\lambda )}_{\bf
p}=\hat p +m \,.
$$
For positrons the corresponding formulae read
$$
v^{(\lambda)}_{\bf p}=-2\lambda i
{ \sqrt{E-m}\,(\mbox{\boldmath $ \sigma$} {\bf n})\,
w_{\bf n}^{(-\lambda) }
\choose \sqrt{E+m}\,w_{\bf n}^{(-\lambda)} } \,,
$$
$$
\bar v^{(\lambda)}_{\bf p}v^{(\lambda')}_{\bf p}=-2 m\,\delta
_{\lambda\lambda'} \,, \;\;\;
\sum _\lambda v^{(\lambda)}_{\bf p}\bar
v^{(\lambda)}_{\bf p}=\hat p-m \,.
$$

In the paper we also use the short notations
$u_j = u^{(\lambda_j)}_{{\bf p}_j}$,
$v_j = v^{(\lambda_j)}_{{\bf p}_j}$ and
$w_{\lambda_j} = w^{(\lambda_j )}_{{\bf n}_j}$.
For the initial electron with momentum ${\bf p}_1$ along
the $z$-axis we put $\theta = \varphi =0$, for the initial
positron with ${\bf p}_2$ opposite to the $z$-axis
$\theta = \pi\,, \; \varphi =0$. For the final positron
with momentum ${\bf p}_4$  $\theta =\pi -\theta_4$ and
$\varphi = \varphi_4$ are used.

\end{document}